\documentstyle[11pt,twoside,epsfig]{article}    
\newcommand{\beq}{\begin{equation}}      
\newcommand{\eeq}{\end{equation}}

\newlength{\dinwidth}  
\newlength{\dinmargin}          
\begin{document}      

%===============================titel ======================================    
\begin{titlepage}
\begin{flushright}      
{\tt LPNHE 96-02}   \\
{\tt March 1996} \\          
\end{flushright}      
\vspace*{1cm}
\begin{center}
\begin{Large}
%\title
\bf{ LOW  \ $Q^2$ \ STRUCTURE FUNCTION  MEASUREMENTS}
\end{Large}  
\vspace*{0.7cm}        

\begin{large}          
%\author
{Gregorio Bernardi}\\          
H1 Collaboration \\
%        on behalf of H1 and ZEUS collaborations\\    
 LPNHE-Paris, 4 place Jussieu, 75252 Paris Cedex 05, France     
\end{large}
\end{center}
%\maketitle        

%\title{ LOW  \ $Q^2$ \ STRUCTURE FUNCTION  MEASUREMENTS}% AT HERA  }  
          
%\author{Gregorio Bernardi\\          
%         H1 Collaboration\\    
% LPNHE-Paris, 4 place Jussieu, 75252 Paris Cedex 05, France}     
%\date{ }  

%\maketitle        
\vspace*{0.3cm}

%================================abstract===============================   
\begin{abstract}  
New results
{\footnote{Talk given at the 2$^{nd}$ Rencontres du Vietnam,
held in H\^o Chi Min City in October 1995. 
The results presented here are extracted from the recent publications of the
the H1 and ZEUS  collaborations~\cite{ZEUSF294,H1F294}.}}
on the   measurement 
of the proton structure function $F_2(x,Q^2)$ are reported        
for momentum transfers squared $Q^2 \geq$ 1.5~GeV$^2$  
 and  Bjorken $x \geq $3.5$\cdot 10^{-5}$,   
using data collected by the HERA experiments H1 and ZEUS in 1994.       
$F_2$ increases significantly with decreasing $x$,   
even in the lowest reachable $Q^2$ region.   
The data are well described by a Next to Leading Order QCD fit,   
and support within the present precision that the rise at low $x$       
within this $Q^2$ range is generated  via the DGLAP        
evolution equations. A comparison with models based on  pomeron 
exchange is also presented. The gluon density is extracted and observed
to rise at low $x$.
\end{abstract}    

%=============================introduction==============================   
          
\section{Introduction}    
The HERA $ep$ collider has been designed to study Deep Inelastic Scattering
(DIS) at very high $Q^2$ where substructure of quarks could be observed.
However in the first 3 years of data operation, which allowed a steady growth
towards the design luminosity of the machine,
 most of the interest has focused on the study
of low $x$, low $Q^2$ DIS, where new tests of perturbative QCD can be 
performed. 
The first observations on the 1992 data showed a rise of the proton      
structure function $F_2(x,Q^2)$ at low $x < 10^{-2}$ with decreasing    
$x$~\cite{H1F292,ZEUSF292}, which was     
confirmed with the more precise data of 1993     
\cite{H1F293,ZEUSF293}.   
Such a behaviour is qualitatively expected in the double leading log    
limit of Quantum Chromodynamics~\cite{ALVARO}. It is, however, not      
clarified 
%%%whether the conventional  DGLAP QCD evolution equation~\cite{DGLAP}
%%%is sufficient to  the rise at low $x$, or if it must be
%%%supplemented by the BFKL  
%%%evolution~\cite{BFKL}.
whether the linear QCD evolution equations, as the    
conventional DGLAP evolution~\cite{DGLAP} in $\ln Q^2$ and/or the BFKL  
evolution~\cite{BFKL} in $\ln(1/x)$, describe the rise of $F_2$ or      
whether there is a significant effect due to 
      nonlinear parton recombination \cite{GLR}.
At low $Q^2$ ($\leq$ 5 GeV$^2$) the new results can be confronted 
to Regge inspired models, which expects a rather flat behaviour as a function
of $x$, in order to study the transition between DIS and photoproduction.
The 1994 data have allowed to reach $Q^2$=1.5 GeV$^2$ and confirm the 
persistance of the rise at low $x$. 
The measurements have been achieved by using dedicated data samples (sect.
 2)
and are discussed and analyzed in terms of perturbative QCD in sect. 3.
          
\section{Structure Function Measurement}    
In 1994 both experiments have reduced the minimum $Q^2$ at which they   
could measure $F_2$ using several techniques:        
%For DIS events at low $Q^2$  the electron is scattered under a large    
% angle    
%$\theta_e$ ( the polar angles $\theta$ are defined w.r.t        
%the proton beam direction, termed "forward" region). 
%Therefore the acceptance of electrons in the backward region has to be  
%increased or the incident electron energy to be reduced         
%to go down in $Q^2$. This was realized as follows.   
i) both experiments were able to diminish the region around the backward 
beam pipe in which the electron could not be measured reliably in 93,   
thus increasing the maximum polar angle of the scattered electron (measured
w.r.t. to proton beam direction).       
This large statistic
\end{titlepage}          
sample, taken with the nominal HERA conditions     
has an integrated luminosity  of about  3~pb$^{-1}$.
%depending  on the analysis/experiment.  
ii) An integrated luminosity of $\sim$ 
60~nb$^{-1}$ of data was collected for which the     
interaction point was shifted by +62~cm,      
in the forward direction,
resulting in an increase of the electron acceptance         
(so-called  "shifted vertex" data sample
%\footnote
%{
%In H1 the low $Q^2$ region was also covered  by analyzing events which  
%originated from the ``early'' proton satellite bunch, allowing to 
%multiply the luminosity by about a factor of 2 between 3.5 and 6.5 GeV$^2$\cite
%{H1F294}.
%}
).
iii) Both experiments used DIS events which underwent initial state     
photon radiation detected in an appropriate photon tagger       
to measure $F_2$ at lower $Q^2$ (so called "radiative" sample).
The incident electron energy which participate in the hard  
scattering is thus reduced, and so is the $Q^2$.          
The luminosity was determined from the measured cross section of        
the Bethe-Heitler reaction $e^-p \rightarrow e^-p\gamma$, measuring the 
 hard photon bremsstrahlung data only.       
The precision of the luminosity measurement is   
1.5\% (3.9\% for the shifted vertex data).

The kinematic variables   
       of the inclusive scattering process $ep \rightarrow eX$  
can be reconstructed in different ways using measured quantities from   
the hadronic final state and from the scattered electron.       
The choice of the reconstruction method for $Q^2$ and $y$       
determines the size of systematic errors, acceptance and radiative      
corrections. The measurements presented here have been obtained with the 
electron (E) and with the $\Sigma$ methods~\cite{sigma} for which the 
$y$, $Q^2$ and $x$ formulae are %%%%%%%%%%($x$ is obtained from $Q^2=xys$)
\begin{equation}  
  y_e   =1-\frac{E'_e}{E_e} \sin^{2}\frac {\theta_e} {2}        
   \hspace*{.8cm}  
   y_{\Sigma} = \frac{\Sigma}{ \Sigma + E'_e(1-\cos{\theta_e})}         
   \hspace*{.8cm}  
   Q^2_{e,\Sigma} = \frac{E^{'2}_e \sin^2{\theta_e}}{ 1-y_{e,\Sigma}}  
   \hspace*{.8cm}  
   x_{e,\Sigma} = \frac{Q^{2}_{e,\Sigma}}{ s y_{e,\Sigma}}  
\end{equation}    
and $E,p_x,p_y,p_z$ are the four-momentum vector components of each particle,
 $E_e$ is the 
electron beam energy, $s$ the squared center of mass energy of the
collision,  $\Sigma\equiv\sum_h{E_h-p_{z,h}}$ and         
the summation is done over all hadronic final  
state particles neglecting their masses.  
The E method, which is independent of the hadronic final state, 
apart from the requirement that the interaction vertex is reconstructed
using the final state hadrons, has at large $y$ the best resolution
in $x$ and $Q^2$ but needs sizeable radiative corrections.
At low $y$ the E method is not  applied due to the degradation
of the $y_e$ resolution as $ 1/y$.    
The $\Sigma$ method, which has small radiative corrections, relies
mostly on the hadronic measurement which has still an acceptable
resolution at low $y$ values and can be used from very low
to large $y$ values.
H1 measures $F_2$ with the E and the $\Sigma$ method and after   
a complete consistency check, in particular at low $x$,
uses the E method for $y>0.15$      
and the $\Sigma$ method for $y<0.15$. ZEUS measures $F_2$ at low $Q^2$
with the E method.
          
%--------------------------------------------------------------------   
The event selection is similar in the two experiments. Events are       
filtered on-line using calorimetric triggers which request an  
electromagnetic cluster of at least 5~GeV not vetoed by a trigger       
element signing a beam background event. Offline, further       
electron identification criteria are applied (track-cluster link,       
shower shape and radius) and a minimum energy of 8(11)~GeV is   
requested in ZEUS(H1). H1 requests a reconstructed vertex       
within 3$\sigma$ of the expected interaction position, while ZEUS       
requires that the quantity $\delta = \Sigma +E'_e(1-\cos\theta)$        
satisfies 35~GeV $< \delta <$ 65~GeV. If no particle escapes detection, 
$\delta = 2 E =$ 55~GeV,   
so the $\delta$ cut reduces the photoproduction background and the      
size of the radiative corrections. The only significant background      
left after the selection comes from photoproduction in which    
a hadronic shower or a photon fakes an electron. In H1 for instance, 
it has been estimated      
consistently      
both from the data and from Monte Carlo simulation and amounts to       
less than 3\% except in a few bins where it can reach values    
up to 15\%. It is subtracted statistically bin by bin and an error of   
30\% is assigned to it.   
          
The acceptance and the response of the detector has been studied and    
understood in great detail by the two experiments: more         
than two millions  Monte Carlo DIS events were generated using   
DJANGO~\cite{DJANGO} and different quark distribution   
parametrizations, corresponding to an integrated luminosity of  
approximately $20$~pb$^{-1}$. The program is based on   
HERACLES~\cite{HERACLES} for the electroweak interaction        
and on  LEPTO~\cite{LEPTO} and ARIADNE~\cite{CDM}  to simulate the hadronic  
final state. HERACLES includes first order radiative corrections,       
the simulation of real bremsstrahlung photons and the longitudinal      
structure function.
For the parton densities, the 
GRV~\cite{GRV} and
the MRS parametrizations~\cite{MRSA} were used.  
The Monte Carlo events, after a detailed simulation based
on 
 the GEANT program,%~\cite{GEANT} 
were subject to the same         
reconstruction and analysis chain as the real data.           

\begin{figure}[tb]       
\begin{center}    
\epsfig{file=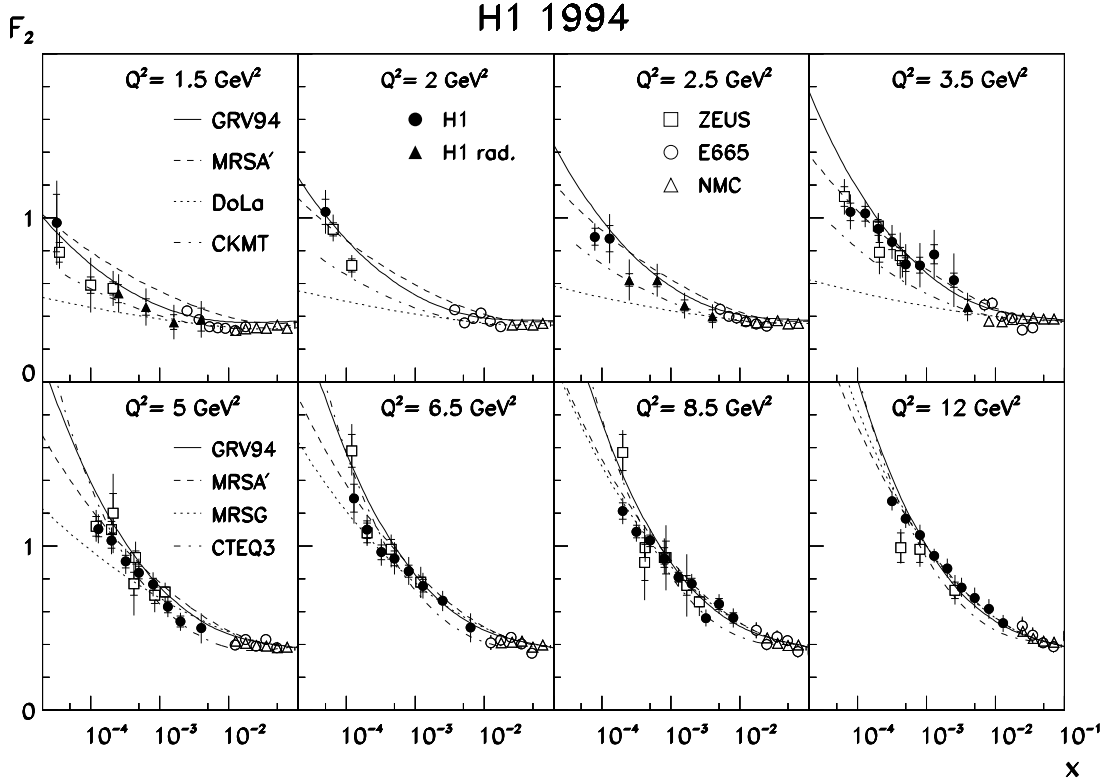,width=14.5cm,       
 bbllx=40pt,bblly=240pt,bburx=540pt,bbury=600pt}      
\end{center}      
\caption[]{\label{lowq2}  
\sl Measurements of the proton structure function $F_2(x,Q^2)$
    in the low $Q^2$ region by H1 and ZEUS are shown together with	  
    the results from the   
    E665 and NMC fixed target experiments. 
    Different  $F_2$ parametrizations are confronted to the    
    data: 
 DOLA and CKMT (only in the upper row of $Q^2$
    bins); CTEQ3M, MRSG and MRSA' (lower row);   
    GRV is shown for the full range.}        
\end{figure}

The  structure function $F_2(x,Q^2)$         
was derived after radiative corrections      
from the one-photon exchange cross section
since effects due to $Z$ boson exchange  are smaller 
than 1\% at low $Q^2$.
   \begin{equation}  
  \frac{d^2\sigma}{dx dQ^2} =\frac{2\pi\alpha^2}{Q^4x}          
    (2-2y+\frac{y^2}{1+R}) F_2(x,Q^2)        
\label{dsigma}    
\end{equation}    
The  ratio $R=F_2/2xF_1 - 1 $      
%has not been measured yet at HERA and        
was calculated using the  QCD relation~\cite{ALTMAR}. 
With  the different data sets available 
%which have different acceptances and use     
%for  a given $Q^2, x$ point different parts of the detectors,
detailed  cross  checks could be made in the kinematic regions of overlap.   
The results were found to be in  very good  agreement   
with each other for all kinematic reconstruction methods used, and
the effect of systematic errors could be monitored:  
%The large available statistics  allows to make       
%very detailed studies on the         
%detector response: efficiencies and calibration. As a result the        
% systematic errors on many effects are reduced, compared to     
%the 1993 data analysis. The effect of   
% the systematic errors can be summarized: 
%refering the       
%reader to the original  $F_2$ publications:
for the E method the main source of error are the energy calibration 
(known at the  1\% level), the knowledge of the electron   
identification efficiency,
%to a lesser extent (except 
%at very low $Q^2$) 
the error on the polar 
angle of the scattered electron, ($\delta\theta=$1mrad) 
and the radiative corrections at         
low $x$.  
For the $\Sigma$ method, the knowledge of the absolute energy scale     
for the hadrons, the fraction of hadrons which stay undetected  
in particular at low $x$, due to calorimetric thresholds        
and to a lesser extent the electron energy calibration are the  
dominating factors.  Further uncertainties common to all methods   
(selection, structure function dependence etc.) were also taken into account.
The total $F_2$ 
errors on the 1994 data ranges between 5 to 10\% in the  10-100 GeV$^2$ 
range and between 10 to 20\% below 10 GeV$^2$.      
The final results on the 1994 data
 of H1 and ZEUS are shown in fig.~\ref{lowq2}
in the new kinematic domain reached
using the radiative and the shifted 
vertex data. 
Compared to the 1993 data analyses
the $F_2$ measurement has been extended 
to lower $x$
(from $1.8 \cdot 10^{-4}$ to   
$3.5 \cdot 10^{-5}$)   and     
$Q^2$ (from $4.5$~GeV$^2$ to  $1.5$~GeV$^2$).        
Both experiments are in good agreement and show that the $F_2$ rise at low
$x$ persist, albeit less strongly, down to the lowest measured $Q^2$=1.5 
GeV$^2$. 
%This rise cannot be attributed to the presence of         
%"diffractive" events in the DIS sample since their proportion has       
%been shown to stay essentially constant (~10\%) independently   
%of $x$ and $Q^2$ \cite{H1DIFF,ZEUSDIFF}.     
A smooth transition to the fixed target experiments E665~\cite{E665} and
NMC~\cite{NMC}
 is observed with
the low $y$ results of H1, allowing to confront all these results to
theoretical expectations.

\section{Low $Q^2$ and Perturbative QCD}     
In fig.~\ref{lowq2} 
are also shown the extrapolations of the $F_2$ 
parametrizations based on some theoretical model 
fitted to the  previous data.  They can be divided in two categories:  one, 
motivated by Regge theory, assumes a pomeron exchange as a dynamical
basis and successfully describes the behaviour of the total cross-sections  
of photoproduction and hadron-hadron collisions;  the other is based on
perturbative QCD and is known to well describe the DIS regime, but
is expected to break down for a given $x$ at some low $Q^2$.
The Regge models were expected to work at least at low $Q^2$, but the
DOLA parametrization which uses a ``soft'' pomeron 
(intercept$\simeq$ 1.08)~\cite{DOLA} largely underestimate $F_2$ at 
low $x$ even at 1.5 GeV$^2$.      
The CKMT model~\cite{CKMT}, which    
assumes that in the present  $Q^2$ range
the "bare" pomeron becomes visible and
has a higher     
trajectory intercept ($\simeq 1.24$),
predicts a weaker rise at low $x$  than observed, except maybe at
 1.5 GeV$^2$.  
% than the "effective" pomeron involved in   
%"soft" interactions ($\sim 0.08$) 
%undershoot the data in a less critica 
%manner. 
These comparisons underline the difference between the behaviour of the total
cross-section of real and virtual photons, since in the HERA kinematic
domain and using the Hand~\cite{Hand} definition of the photon flux
 $\sigma_{tot}^{\gamma^* p}$ can be expressed as
%\begin{equation}  
$
\sigma_{tot}^{\gamma^* p}(x,Q^2) \simeq \frac{4~\pi^2 \alpha}{Q^2} F_2(x,Q^2). 
$
%\end{equation} 
 
\begin{figure}[tbh]    \unitlength 1mm
\begin{center}
\begin{picture}(147,85)
\put(-11,-3){
\begin{picture}(0,0) \put(10,75){({\large \bf a})} \end{picture}
\begin{picture}(0,0) \put(95,75){({\large \bf b})} \end{picture}
\epsfig{file=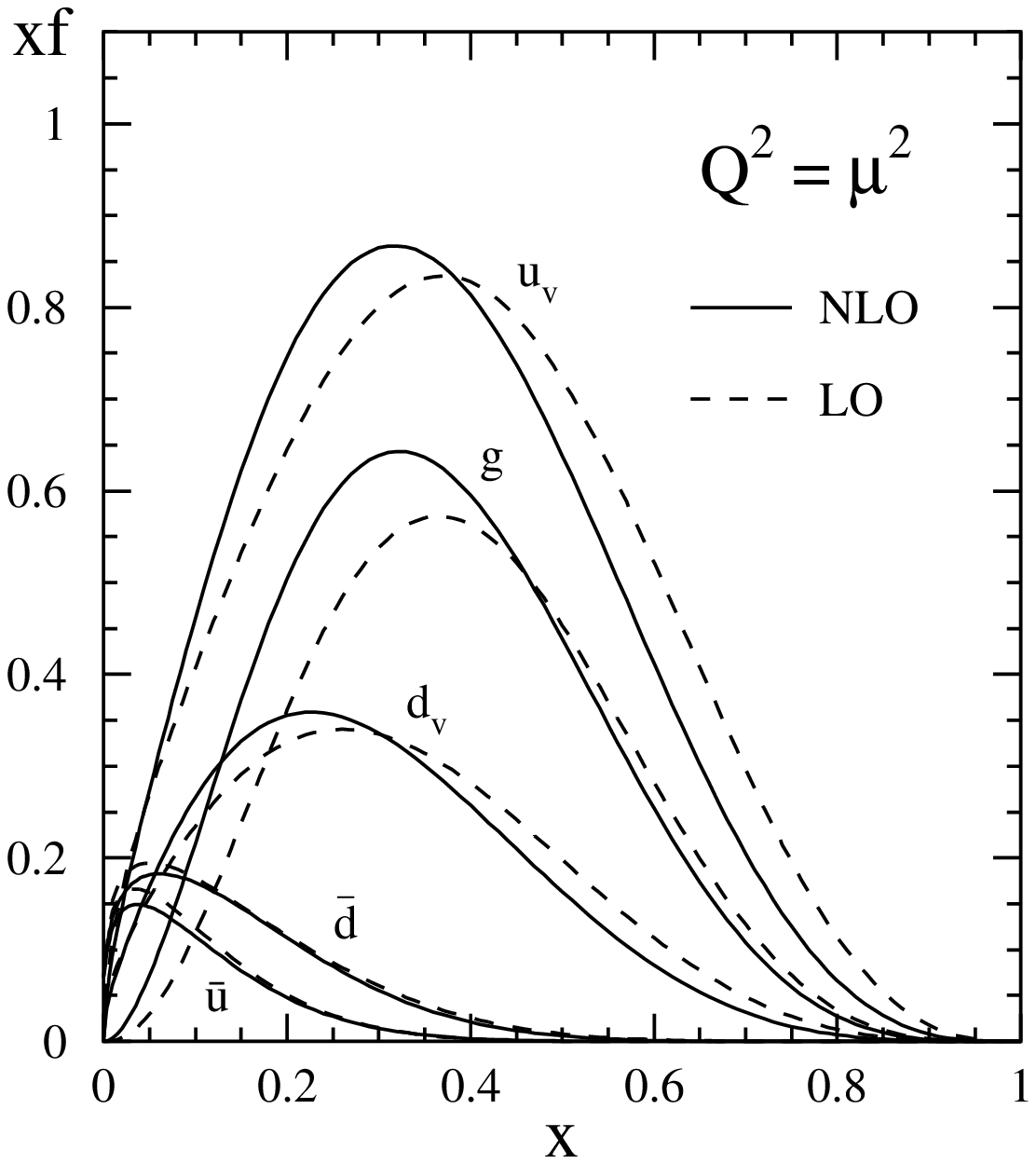, 
width=8.0cm,bbllx=150pt,bblly=260pt,bburx=480pt,bbury=570pt}
\epsfig{file=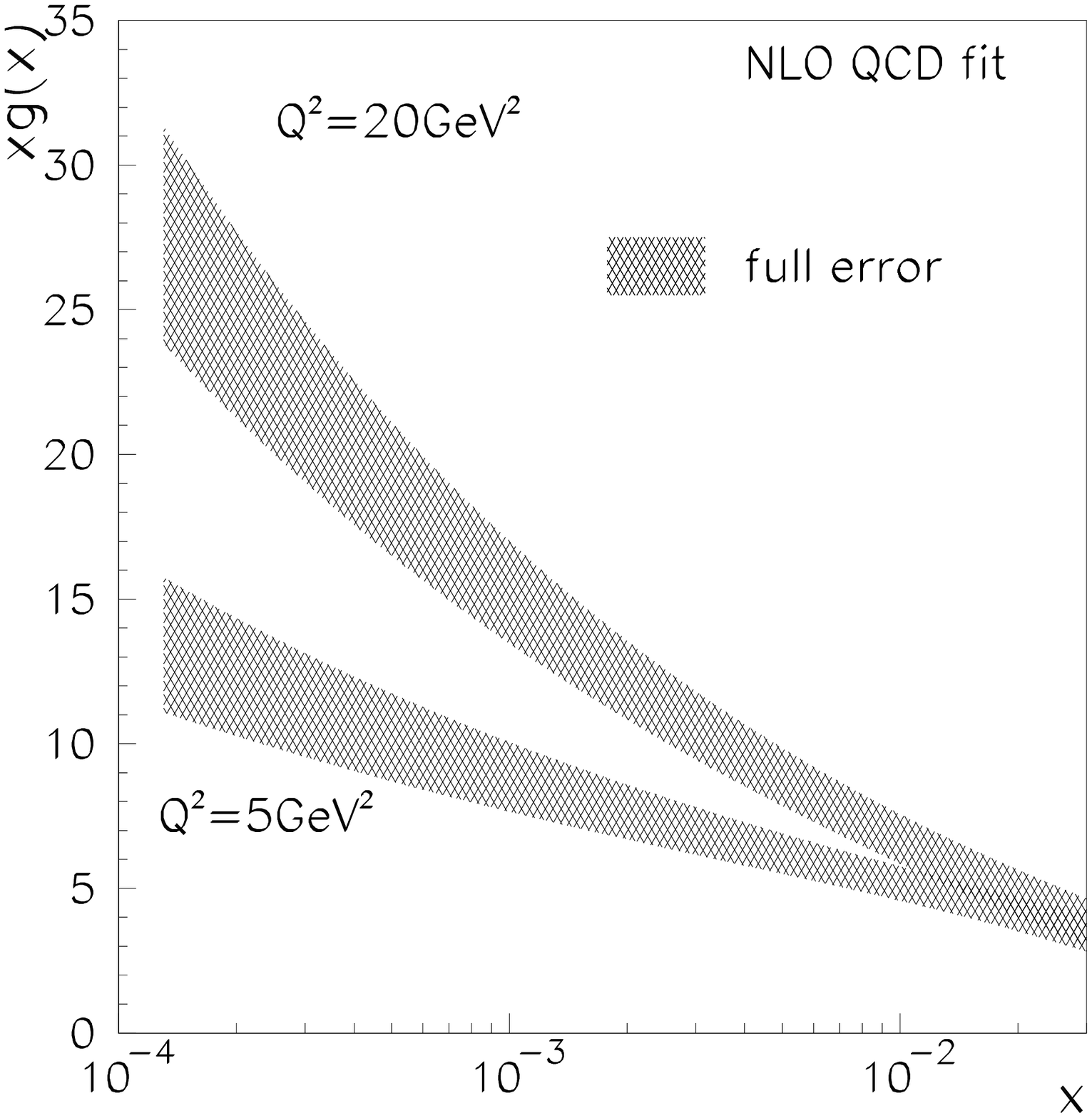,
width=8.3cm,bbllx=10pt,bblly=160pt,bburx=555pt,bbury=710pt}   
}
\end{picture}
\end{center}   
\caption[]{\label{glugrv}      
\sl 
a) Parton densities (valence quarks (u$_v$,d$_v$),
gluon (g) and sea quarks) of the GRV model at the initial energy scale
$\mu^2=0.34$ GeV$^2$.
b) Gluon density at
5 and 20 GeV$^2$ determined by a NLO fit
to the H1,NMC and BCDMS data. The error bands represent the full error
except for the uncertainty on $\Lambda$.}
\end{figure}

The parametrizations based on the DGLAP QCD evolution equations describe the
data remarkably well, as expected above 5-10 GeV$^2$, but surprizingly
at values around 1 or 2 GeV$^2$ where non-perturbative effects were 
believed to distort the DGLAP picture. 
The MRSA' parametrizations of the 
parton densities are defined at $Q^2_0=$ 4 GeV$^2$, then 
evolved in $Q^2$ and fitted to previous experimental data, including the 1993
HERA data. The agreement observed above 10 GeV$^2$ confirms that the 1993 and
1994 HERA results are  compatible. Between 1.5 and 10 GeV$^2$ the good
description  tells us that within the present precision perturbative QCD
can be applied in this range.
More striking is the confirmation of the pre-HERA prediction of the $F_2$
\begin{figure}[hbt]
%\begin{minipage}[t]{7.8cm}  
\begin{center}    
\epsfig{file=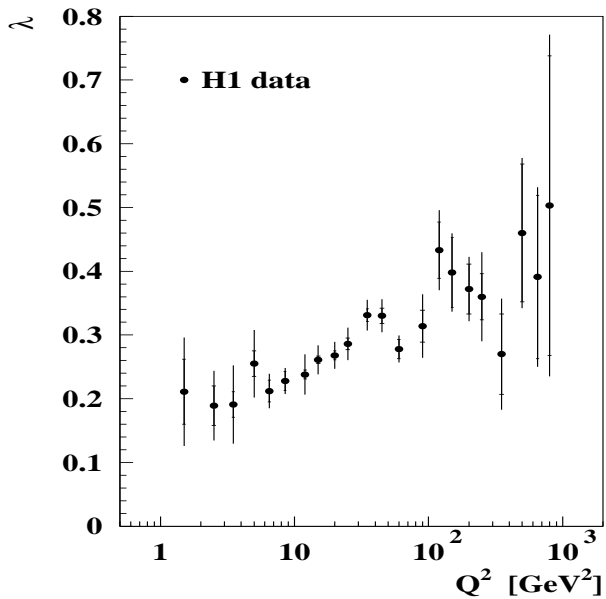,height=8.0cm, 
width=8.0cm,bbllx=80pt,bblly=140pt,bburx=470pt,bbury=660pt}
\end{center}      
\caption[]{\label{expox}  
\sl Variation  of the exponent $\lambda$  from fits
of the form $F_2 \sim x^{-\lambda}$ at fixed $Q^2$
values.} 
%\end{minipage}   \hfill       
\end{figure} 
rise at low $x$ by the
GRV model~\cite{grv1}
 which conjectured at a very low energy scale ($\mu^2$=0.34 GeV$^2$)
that the proton is formed by valence-like partons as shown in fig.~\ref{glugrv}
and that the DGLAP equations can be applied to generate ``radiatively'' the
rise of the gluon and sea-quark density at low $x$, when evolving towards 
higher $Q^2$. 
The H1 and ZEUS results are very well  described by the GRV model at low $Q^2$
as can be seen in fig.~\ref{lowq2} but also in the full HERA kinematic range,
from 1.5 to 5000 GeV$^2$~\cite{moscow}.
This success support the idea that the rise at low $x$ is a direct 
consquence of the DGLAP equations, and that non-perturbative effects
are relatively weak at low $x$ and low $Q^2$.
The evolution with $Q^2$ of the strength of the rise can be quantified
by fitting an $x^{-\lambda}$ (or equivalently a $W^{2\lambda}$, $W$ being
the invariant mass of the $\gamma^{\star}-p$ system) function at fixed $Q^2$
to $F_2(x), x < 0.1$. The values of $\lambda$ obtained by the fit in 
each $Q^2$ bin are displayed in fig~\ref{expox} and clearly confirm 
%\hfill
%\begin{figure}[htb]
%\begin{center}    
%\epsfig{file=be2.ps,width=5cm,height=7cm,       
% bbllx=10pt,bblly=400pt,bburx=300pt,bbury=650pt}      
%\end{center}      
%\caption[]{\label{bemc}  
%\sl Preliminary measurement of the proton structure function $F_2(x,Q^2)$
%    in the low $Q^2$ region by H1 and ZEUS.} 
%\end{figure}  
the long time prediction made %in leading order
for asymptotic free field theories like
QCD~\cite{ALVARO} of a rise of $F_2$ at low $x$, and that
the strength of this rise increases with $Q^2$. 
With the present data, it is however not possible
to know  precisely this strength below 5 GeV$^2$, thereby postponing
a definite test of perturbative QCD in this region. 

\begin{figure}[bth]    \unitlength 1mm
\begin{center}
\begin{picture}(155,100)
\put(-11,-3){
\begin{picture}(0,0) \put(17,0){{\large {\it 10$^{-4}$}}} \end{picture}
\begin{picture}(0,0) \put(28,0){{\large {\it 10$^{-2}$}}} \end{picture}
\begin{picture}(0,0) \put(45,0){{\large {\it 10$^{-4}$}}} \end{picture}
\begin{picture}(0,0) \put(56,0){{\large {\it 10$^{-2}$}}} \end{picture}
\begin{picture}(0,0) \put(73,0){{\large {\it 10$^{-4}$}}} \end{picture}
\begin{picture}(0,0) \put(84,0){{\large {\it 10$^{-2}$}}} \end{picture}
\begin{picture}(0,0) \put(101,0){{\large {\it 10$^{-4}$}}} \end{picture}
\begin{picture}(0,0) \put(112,0){{\large {\it 10$^{-2}$}}} \end{picture}
\begin{picture}(0,0) \put(129,0){{\large {\it 10$^{-4}$}}} \end{picture}
\begin{picture}(0,0) \put(140,0){{\large {\it 10$^{-2}$}}} \end{picture}
\epsfig{file=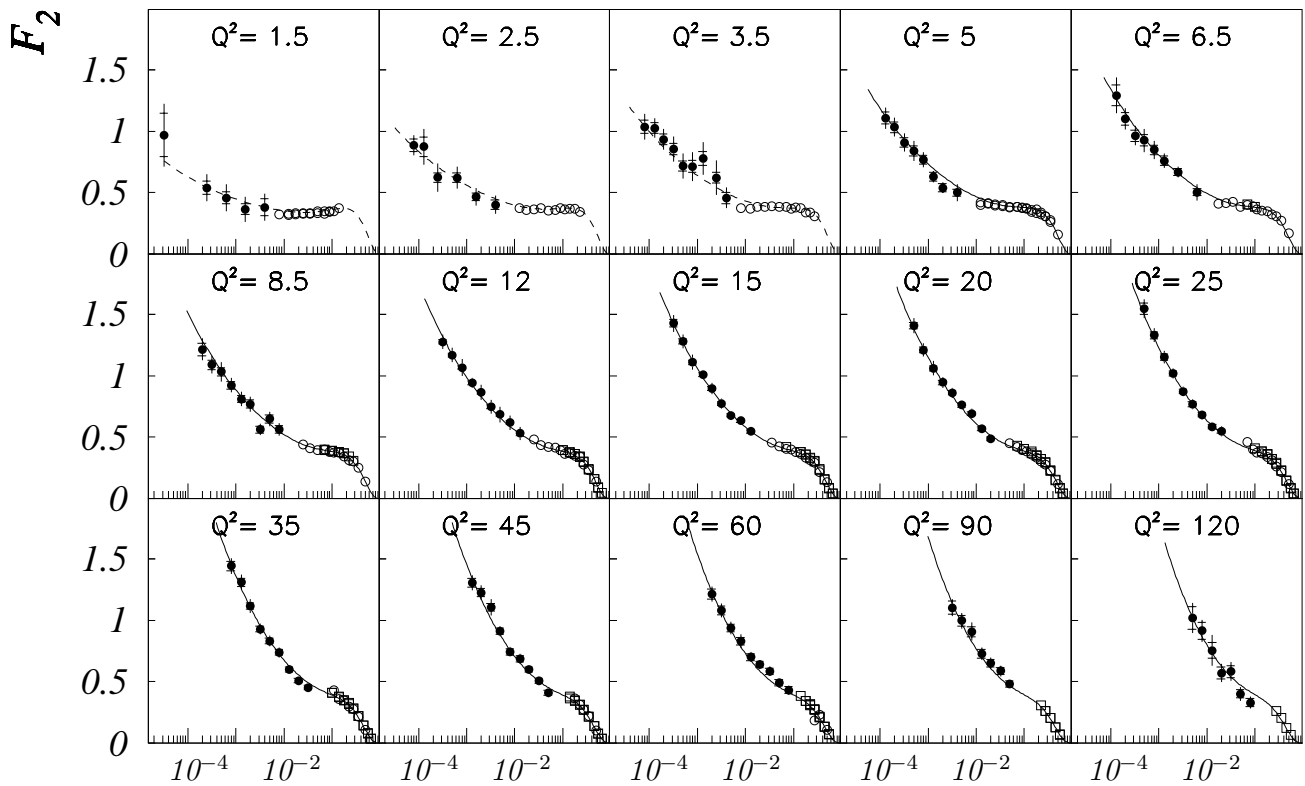,width=15.5cm,%    
 bbllx=85pt,bblly=380pt,bburx=550pt,bbury=690pt} 
}    
\end{picture}
\end{center}      
\caption[]{\label{f2x}    
\sl H1 measurement (black circles) 
    of the proton structure function $F_2(x,Q^2)$   
    as function of $x$ in different bins of $Q^2$. The          
    inner error bar is the statistical error. The full error    
    represents the statistical and systematic errors added      
    in quadrature. The curve represent a NLO QCD fit to the H1, BCDMS (open
squares) and NMC (open circles) data at Q$^2 > $5 GeV$^2$.}       
\end{figure}              

To make fully use of the new precision reached with the 1994 data, the H1
collaboration has performed 
%\footnote{The ZEUS collaboration is soon to release the same kind
%of analysis}
a Next-to-Leading Order (NLO) QCD fit on the H1, BCDMS and NMC
data with the conditions
$Q^2 >$ 5 GeV$^2$, and $x <$ 0.5 if $Q^2 <$ 15 GeV$^2$ to avoid 
higher-twists effects.
The H1 measurements which extend up to  5000 GeV$^2$ (they are shown
up to 120 GeV$^2$ in fig.~\ref{f2x}) were fitted successfully and
allow to constrain the gluon density at low $x$. The parton densities were
parametrized at $Q^2_0$=5 GeV$^2$, in particular the gluon was expressed
with 3 parameters as $xg(x)= A_gx^{B_g}(1-x)^{C_g}$.
The quark and antiquark components of the sea were assumed to be equal,
and $\bar{u}$ set equal to $\bar{d}$. As determined in ~\cite{ccfr},
the strange quark density was taken to be $\bar{s}=(\bar{u}+\bar{d})/4$.
Further constraints were coming from the quark counting rules and the
momentum sum rules. For $\Lambda$ the value of 263 MeV was taken~\cite{qcdbcd}.
A detailed treatment of the $F_2$ errors propagation on the 
gluon density has been done, resulting in the error bands of fig~\ref{glugrv}b
which represent $xg(x)$ at 5 and 20 GeV$^2$. 
A variation of 
$\Lambda$ by  65 MeV 
gives a  change of 9\%  on the gluon density 
at 20 GeV$^2$ which has not been added to the error bands.
The accuracy  of this determination of $xg$  
is better by  about a factor of two  than  the H1 result based on
the 1993 data~\cite{qcdfit}.
   
A  rise of the gluon density towards low $x$ is observed
which  is related  to the behaviour of $F_2 \propto x^{-\lambda}$.
Accordingly, the rise of $xg$ towards low $x$ increases with increasing $Q^2$.
Finally we can observe in fig.~\ref{f2x} 
that the data at 
$Q^2 < 5 $~GeV$^2$, which were excluded from the fit, are still
well reproduced by the fit evolved backwards in $Q^2$.
More data at low $x$ and $Q^2 < 1
 $~GeV$^2$ are nevertheless needed to be able
to test the hypothesis of a gluon density which would take the valence-like
shape displayed in fig.~\ref{glugrv}a when $Q^2 \rightarrow$ 0.3 GeV$^2$,
 and more generally, to better understand the dynamics at low $Q^2$ and high
parton densities.
The HERA experiments, which have last year upgraded the capabilities of
their backward detectors,  will be able to reach these low 
$Q^2$ with the data taken in 1995 and 1996. 
           
\vspace*{0.3cm}
\noindent
{\Large{{\bf Acknowledgements}}}    
%=====================    

\normalsize       
\vspace*{0.1cm}
\noindent         
I would like to thank the organizers and in particular 
Kim and J. Tr\^an Thanh V\^an
to have realized  such an interesting and fruitful
exchange workshop in this fascinating country.  
I would also like to thank my close collaborators,
Ursula Bassler, Beatriz Gonzalez-Pineiro and Fabian Zomer,
all the friends of the
H1 structure function group and the H1 and ZEUS collaboration
with whom we obtained the results
described above. 
Special thanks also to Marie for her constant help and support, and for sharing
our unforgettable vietnamese journey.
%Special thanks go to Ursula             
%for her  help in the finalization of this         
%paper.   

%---------------------------------------------------------------------- 
     
\end{document}